\documentstyle[amssymb,preprint,aps]{revtex}
\tightenlines
\begin{document}
\draft

\title{Metal-insulator transition in amorphous alloys}

\author{A.~M\"obius$^1$ and C.J.~Adkins$^2$}

\address{
$^1$Institute for Solid State and Materials Research Dresden,
D-01171 Dresden, Germany, e-mail: a.moebius@ifw-dresden.de,\\
$^2$Cavendish Laboratory, Madingley Road, Cambridge CB3 0HE, UK,
e-mail: cja13@cam.ac.uk}

\maketitle

\begin{abstract}
We focus on the central problem of discriminating between metallic and
insulating behaviour in amorphous alloys formed between a semiconductor 
and a metal. For this, the logarithmic temperature derivative of the 
conductivity, $w = \mbox{d} \ln \sigma / \mbox{d} \ln T$, has proved
over recent years to be very helpful in determining the critical value
$x_{\rm c}$ of the metal content $x$ for the metal-insulator transition
(MIT). We show that, for various amorphous alloys, recent experimental 
results on $w(T,x)$ are qualitatively inconsistent with the usual 
assumptions of continuity of the MIT at $T = 0$ and of 
$\sigma(T,x_{\rm c})$ being proportional to a power of $T$. These 
results suggest that $w(T,x_{\rm c})$ tends to 0 as $T \rightarrow 0$, 
in which case the MIT should be discontinuous at $T = 0$ (but only 
there), in agreement with Mott's hypothesis of a finite minimum metallic 
conductivity.
\end{abstract}


\noindent
{\bf Abbreviations}\\
EDX: energy dispersive X-ray analysis\\
MIT: metal-insulator transition\\
MOSFET: metal oxide semiconductor field effect transistor\\
RBS: Rutherford backscattering analysis

\begin{center} \noindent
{\bf Introduction}
\end{center}

Binary amorphous alloys, composed of a semiconductor and a metal, belong 
to a broad class of materials which exhibit a transition between 
metallic and activated behaviour of the electrical conductivity caused 
by the variation of chemical composition, temperature, stress or 
magnetic field. Other such systems are granular metals, heavily doped 
crystalline semiconductors, conducting oxides, both amorphous and 
crystalline, including the high-temperature superconductors, 
quasicrystals, and doped polymers. Disorder plays an important role in 
many of these materials. This may involve the structure of an amorphous 
alloy or the random positions of dopant atoms in an ordered matrix. 
Simultaneously, various transport properties, in particular the 
temperature dependence of the conductivity, exhibit the same 
characteristic qualitative features in many of these substances.

The behaviour of systems that undergo a metal-insulator transition (MIT)
has attracted much interest, theoretical as well as experimental, for 
more than forty years. Milestones were the concepts of Anderson 
localisation \cite{A58}, variable-range hopping \cite{Mott.68}, minimum 
metallic conductivity \cite{Mott.72}, the Coulomb glass \cite{P70,ES75}, 
the scaling theory of localisation of noninteracting electrons 
\cite{AALR79}, and the renormalisation group approach incorporating the 
electron-electron interaction into localisation theory \cite{F83,F84}. 
In the literature, many surveys have appeared in this wide area 
\cite{MD79,SE84,EP85,LR85,M85,Mott.90,F90,KM93,BK94,E95,L98}.

Nevertheless, despite all these efforts, our knowledge about electrical
transport in disordered systems in the vicinity of the metal-insulator 
transition is still rather uncertain. This is strikingly illustrated by 
two examples: (i) The scaling theory of localisation of noninteracting 
electrons \cite{AALR79} denies the existence of a MIT in two-dimensional 
systems. This result was regarded almost as a dogma for many years. 
Thus, the discovery of a MIT in MOSFETs with very high electron mobility 
by Kravchenko et al.\ \cite{KK94} in 1994 came as a big surprise. (ii) 
For three-dimensional systems, the value of the critical exponent of the 
zero-temperature conductivity has been a matter of controversial debate 
since its first measurements, see e.g.\ \cite{PR82,HB83,HT88,SH93}. So 
far, there seems to be no generally accepted solution of this 
`critical-exponent puzzle'. Certainly, one of the reasons why this 
question has not yet been solved, is the uncertainty in identifying the 
critical concentration in such experiments (see references \cite{M89} 
and \cite{C94} commenting on the related publications \cite{HT88} and 
\cite{SH93}, respectively).

Among the amorphous systems with a MIT which have been investigated for 
the last twenty years, metastable alloys of semiconductors with noble 
metals, transition metals, and rare earths play the most important role. 
They can be produced by various methods such as thermal and 
electron-beam evaporation, sputtering, laser quenching, and ball 
milling. This subject has been investigated so intensively that it is 
impossible to give here a complete survey of all related studies of 
structure and transport properties of amorphous alloys in the vicinity 
of the MIT. Thus the following selection, listed according to the alloys 
investigated, contains merely the more important publications on 
low-temperature transport and, in some cases, structure:
Ge-Al \cite{LD85}, Ge-Au \cite{DM81,MM81,EF91}, Ge-Co \cite{AM96},
Ge-Cu \cite{AG86}, Ge-Cr \cite{AM96,EG91,HV93,BE94}, 
Ge-Fe \cite{AM96,AM93,AM94,RR94,AM95,RB95}, Ge-Mn \cite{AM96}, 
Ge-Mo \cite{RB95,YG88}, Ge-Ni \cite{AM96}, 
Si-Au \cite{M80,NY82,MD83,AK89,AC90,FL93},
Si-Cr \cite{BE94,ME83,MV85,M87,M90a,M90b,OS89,M90c,SB91,VH92,HS93},
Si-Fe \cite{RR94,DY89,FR95}, Si-Gd \cite{ST86,XZ99},          
Si-Mn \cite{DY89,YW95,YD97,YD98}, Si-Nb \cite{HB83,SH91,CL96,HC98},
Si-Ni \cite{EF89,DB89,BA91,AB91,AA92,DA93,BS94,RH97a,RH97b,IY97,MF98},
Si-Pd \cite{TF95}, Si-Pt \cite{LE87}, Si-Re \cite{VH95},
Si-Ta \cite{WP96,WP98}, 
Si-Tb \cite{ST86}, Si-V \cite{OR87,BH90,MI97}, Si-Y \cite{ST88,ST90}.
The transport studies from this list are not only devoted to 
conductivity, in particular its temperature dependence, but in many 
cases also to magnetoresistance; several concern the thermopower too. 
Moreover, the formation of a Coulomb gap has been studied in various 
alloys by tunnelling experiments. The structural investigations, e.g.\ 
of Si-Fe \cite{RR94,FR95} and Si-Ni \cite{EF89}, have demonstrated that 
the concept of a homogenous amorphous substance is an idealisation. In 
reality, small ($\sim$ 1~nm) clusters with enlarged metal content, with 
composition close to an intermetallic phase, can be formed. Thus a 
comparison with related granular systems would be interesting. In 
particular for Ge-Al, granular films have been studied in detail
\cite{MR93,ER93,RS94}. Besides the substances listed above, several 
other amorphous materials exhibit a MIT. However, here we refer for 
comparison only to amorphous InO$_x$ \cite{O86,R91,GA98}.

The amorphous semiconductor-metal alloys resemble each other concerning 
the dependence of the conductivity $\sigma$ on temperature $T$ and metal 
content $x$. The MIT very often occurs for an $x$ value between 0.1 and 
0.2. Besides superconductivity in some cases, the only qualitative 
feature by which these materials may be distinguished from each other 
seems to be the occurrence of a resistivity dip at low $T$ in the 
metallic region, observed in some of the alloys only \cite{AM96}. 
However, when preparing the same alloy by different technologies, 
considerable quantitative differences concerning $\sigma(T,x)$ can be 
present. The typical behaviour of $\sigma(T,x)$, as well as the 
influence of preparation conditions are represented in Fig.\ 1 for the 
system Si-Ni. In trying to understand the differences between the three 
kinds of samples shown, it is not sufficient only to consider the 
different amounts of impurities built in from the residual gas, but one 
has also to take into account structural aspects -- note that aging and 
annealing can considerably change $\sigma(T)$, see e.g.\ 
\cite{DM81,MV85,DY89,BS94}. Therefore, it can be quite problematic to 
apply theories developed for homogeneous systems to the description of 
the conduction processes in such alloys. However, due to diverging 
correlation lengths as the MIT is approached, we can nevertheless hope 
that there are some universally valid relations in this limit.

The above mentioned uncertainties in the interpretation of experiments 
by means of current theory, as well as the variability of the structure 
of seemingly amorphous films suggest that, in interpreting data, one 
should look for an approach that is as unbiased as possible. It is an 
important result of recent years that, even on a phenomenological level,
such an approach can lead by means of comparatively simple mathematical 
considerations to interesting conclusions on the character of the MIT.
In the next section, we focus on this aspect, analysing the general 
problems of discriminating between metallic and insulating samples, and 
demonstrating how useful the logarithmic $T$ derivative of $\sigma$ can 
be for this purpose. Then, on the basis of experimental data for various 
alloys, we shall show that Mott's hypothesis on the existence of a 
finite minimum metallic conductivity may well be correct after all, at 
least for certain three-dimensional systems. (For this reason, we shall 
not discuss fitting of formulae which result from microscopic theories 
yielding a continuous transition.) In a fourth section, devoted 
specifically to amorphous Si$_{1-x}$Cr$_x$, we explain a 
phenomenological model derived for this substance through searching for 
universal features in $\sigma(T,x)$, in particular performing a scaling 
analysis. Finally, after summarising briefly, we list for future work a 
series of questions which are implied by this review.

\begin{center} \noindent
{\bf Basic considerations}
\end{center}

Consider the electric transport in a (three-dimensional) amorphous 
alloy as a function of temperature $T$ and a control parameter $x$,
describing the chemical composition. Assume, this alloy undergoes a MIT 
when $x$ is varied, where $\sigma$ increases with $x$. Each study of 
this MIT needs first of all clear definitions of the terms metal and 
insulator. To relate them to any finite values of conductivity, 
temperature and frequency is necessarily an arbitrary judgement. Thus 
the only unambiguous discrimination between metallic and insulating 
samples is based on the zero-temperature limit. If the direct current 
conductivity $\sigma$ tends to some finite value, then the sample is 
regarded as metallic; if, however, $\sigma \rightarrow 0$ as 
$T \rightarrow 0$, it is insulating. This definition is simple, but it 
cannot be utilised directly since each measurement has some finite 
lowest experimentally accessible temperature $T_{\rm lea}$. Hence, one 
always depends on an extrapolation. It can be based on a generalisation 
of experimental experience or on a microscopic theory. Moreover, we 
can only assume that the physical processes which we have studied within 
some region of finite $T$ remain the relevant ones down to arbitrarily 
low $T$.

In this sense one usually considers a sample with an exponential
$T$ dependence of $\sigma$, 
\begin{equation}
\sigma(T,x) = \sigma_1 \, \exp\{-[T_0(x)/T]^\nu\}\,,
\label{exp_hopping}
\end{equation}
as insulating at $T = 0$. Here, the exponent $\nu$ often has one
of the values $1/4$ and $1/2$ according to variable-range hopping
theories by Mott \cite{Mott.68} and by Efros and Shklovskii \cite{ES75}, 
who ignore or take into account electron-electron interaction, 
respectively; in other cases, $\nu$ is determined by fitting. The degree 
of reliability of this extrapolation approach depends on the width of 
the $\sigma$ interval considered. It should enclose several orders of 
magnitude so that algebraic corrections of Eq.~(\ref{exp_hopping}) can 
be neglected. Thus $T_{\rm lea}$ must be considerably smaller than 
$T_0$.

On the other hand, samples are considered as metallic if the relation
\begin{equation}
\sigma(T,x) = a(x) + b(x) \cdot T^p
\label{met_extrap}
\end{equation}
with $a > 0$ holds. Mostly, such analyses are performed for $p = 1/2$ 
or $1/3$ resulting from theories by Altshuler and Aronov 
\cite{AA79}, and Newson and Pepper \cite{NP86}, respectively. The former 
theory models the superposition of electron-electron interaction and 
disorder, but it is a perturbation theory so that its applicability 
close to the transition is at least questionable. The latter theory 
considers the $T$-dependent drop of the diffusion constant as the 
decisive variation, and yields a power law with exponent $1/3$ for 
$x = x_{\rm c}$ where $a = 0$.

The extrapolation is particularly difficult if a strong, but
non-exponential $T$ dependence of $\sigma$ is observed. In such a case,
two interpretations are possible: (i) The sample could be metallic, 
provided there is a `good' fit to Eq.~(\ref{met_extrap}) with 
$a > 0$. (ii) The sample could exhibit activated conduction with a
small characteristic temperature $T_0$ being at most of the order of 
$T_{\rm lea}$. If, as is generally taken for granted, the parameter 
$T_0$ vanishes continuously as the MIT is approached from the insulating 
side, there is always a finite $x$ interval where the situation (ii) is 
realized, as visualised by Fig.\ 2. But, to the best of our knowledge, 
an appropriate microscopic theory for a quantitative analysis within 
this non-exponential $(x,T)$ region is still missing.

Thus, often the samples with an `intermediate strength of the $T$ 
dependence' are only analysed in terms of (i), and the alternate 
possibility (ii) is ignored. However, what one can easily do, and should 
always do, is to perform a consistency check of the description by 
Eq.~(\ref{met_extrap}). For that, we consider the logarithmic 
temperature derivative of the conductivity,
\begin{equation}
w(T) = \mbox{d} \ln \sigma(T) / \mbox{d} \ln T\,.
\label{log_deriv}
\end{equation}
Its behaviour is qualitatively different on the two sides of the MIT.
For insulating samples, Eq.~(\ref{exp_hopping}) yields,
\begin{equation}
w(T)= \nu\,(T_0/T)^\nu\,,
\end{equation}
so that $w(T \rightarrow 0) = \infty$ for $x < x_{\rm c}$. On the other 
hand, if the conduction is metallic, we get from Eq.~(\ref{met_extrap})
\begin{equation}
w(T)= p\,b\,T^p / (a + b\,T^p)\,.
\end{equation}
Hence, $w(T \rightarrow 0) = 0$ for $x > x_{\rm c}$. If this limiting 
behaviour is not obvious from the experimental data, the validity of 
Eq.~(\ref{met_extrap}) should be questioned. This consistency check was 
first used for crystalline Si:(P,B), where it showed that the 
discrimination between insulating and metallic samples, based only on 
(i), is not satisfactory \cite{M89,HT89}. It proved to be helpful also 
in several other experiments \cite{YD98,RH97a,RS94,GA98,FZ96}.

Note that the consideration of $w$ can even permit a classification, 
when a sample belongs to the $x$ region where the activated character 
cannot be detected via exponential $\sigma(T)$, i.e.\ where 
$w \stackrel{_{\textstyle <}}{_{\textstyle \sim}} \nu$. Assume, for a 
certain sample, we have observed that $\mbox{d} w / \mbox{d} T \le 0$ at 
$T_{\rm lea}$. Experimental experience (see next section) seems to 
suggest that, once $w$ has begun to increase with decreasing $T$, it 
continues to do so as $T$ tends to 0. If, therefore, we extrapolate 
merely the validity of $\mbox{d} w / \mbox{d} T \le 0$ to arbitrarily 
low $T$, we reach the conclusion that, as $T \rightarrow 0$, $\sigma$ 
vanishes at least as fast as some (positive) power of $T$. Thus the 
observation of $\mbox{d} w / \mbox{d} T \le 0$ at $T_{\rm lea}$ suggests 
strongly that the sample is an insulator at $T = 0$.

In this way, the calculation of $w$ allows us to check whether small 
deviations of the experimental data from the fit of 
Eq.~(\ref{met_extrap}) are understandable as random scatter of the data 
points, or as some small correction, or whether they indicate a 
qualitative change to be expected if lower temperatures were accessible.
Inspection of $w$ also provides a sensible accuracy test concerning 
thermometry problems and thermal decoupling. Sudden changes of $w$ 
occurring for all samples at roughly the same $T$ value, for example, 
should be investigated with care, cf.\ \cite{M90c}.

However, the analysis of $w$ cannot only be used in classifying a given
sample as metallic or activated. This quantity contains still more 
information \cite{MF98}: A fingerprint of the character of the MIT at 
$T = 0$ can be contained in the graph of $w(T,x={\rm const.})$ for a set 
of samples measured at finite $T$. To illustrate this, we calculate 
$w(T,x)$ for two simple, qualitative models. On the basis of `usual' $T$ 
dependences, they are constructed so that continuity of $\sigma$ at 
$x_{\rm c}$ for arbitrary finite $T$ and monotonicity of 
$\sigma(T = {\rm const.},x)$ are guaranteed.

First we assume the transition to be continuous also at $T = 0$:
\begin{equation}
\sigma(T,x) = \left\{ \begin{array}{l}
T^{1/2} \exp\{-[T_0(x)/T]^{1/2}\}\\ a(x) + T^{1/2}
\end{array} \quad\quad\mbox{for}\quad\quad
\begin{array}{l} x < x_{\rm c} \\ x \ge x_{\rm c} \end{array} \,,
\right.
\label{cont_model}
\end{equation}
where $T_0(x \rightarrow x_{\rm c} - 0) = 0$, and
$a(x \rightarrow x_{\rm c} + 0) = 0$. For simplicity, all quantities
are dimensionless. Corresponding $w(T)$ curves are presented in Fig.\
3: The curve, which separates activated and metallic regions, is a 
straight line parallel to the $T$ axis, $w(T,x_{\rm c}) = p = 1/2$. 
There are no pieces of `insulating' $w(T)$ curves with $w < p$. Note 
that, if $w < p$, always $\mbox{d}w/\mbox{d}T > 0$.

Next we assume the transition to be discontinuous, but only at
$T = 0$:
\begin{equation}
\sigma(T,x) = \left\{ \begin{array}{l}
(1 + T^{1/2}) \exp\{-[T_0(x)/T]^{1/2}\}\\ a(x) + T^{1/2}
\end{array} \quad\quad\mbox{for}\quad\quad
\begin{array}{l} x < x_{\rm c} \\ x \ge x_{\rm c} \end{array} \,,
\right.
\label{disc_model}
\end{equation}
where $T_0(x \rightarrow x_{\rm c} - 0) = 0$, and
$a(x \rightarrow x_{\rm c} + 0) = 1$. Fig.\ 4 shows corresponding $w(T)$ 
curves. In contrast with Fig.\ 3, in the activated region, this graph 
exhibits minima of $w(T,x = {\rm const.})$, where the values of 
$w_{\rm min}(x)$ become much smaller than $p$ as the MIT is approached. 
This feature arises from the limiting behaviour of $w_{\rm min}(x)$ and 
of the related $T_{\rm min}(x)$ as $x \rightarrow x_{\rm c}-0$; both 
these quantities tend to zero. We would like to emphasise that, 
therefore, Fig. 4 exhibits also such `insulating' $w(T)$ curves which 
have low-$T$ pieces where simultaneously $\mbox{d} w / \mbox{d} T < 0$ 
and $w < p = 1/2$ hold. Moreover, 
$w(T,x_{\rm c}) = T^{1/2} / [2 \,(1+T^{1/2})]$ tends to 0 as 
$T \rightarrow 0$.

The correlation between $w(T,x_{\rm c})$ vanishing as 
$T \rightarrow 0$ and $\sigma(0,x)$ jumping at $x_{\rm c}$ is not a 
special feature of the model (\ref{disc_model}) we have used. We can 
demonstrate this by examining the consequences of the limiting behaviour 
of $w$. Assume, at the MIT, $w$ tends to 0 according to some power law 
as $T \rightarrow 0$, 
\begin{equation}
w(T,x_{\rm c}) = c \, T^q\,,
\end{equation}
where $q > 0$. Hence,
\begin{equation}
\sigma(T,x_{\rm c}) = \sigma(0,x_{\rm c})\,\exp(c \, T^q / q)\,.
\end{equation}
The exponential factor is finite for arbitrary $T$ so that obtaining a 
finite value of $\sigma$ at any measuring temperature $T_{\rm m}$ for 
$x = x_{\rm c}$ indicates that $\sigma(0,x_{\rm c})$ is finite as well, 
and that, since $\sigma(0,x) = 0$ for $x < x_{\rm c}$, the function 
$\sigma(0,x)$ must have a discontinuity at $x_{\rm c}$.

\begin{center} \noindent
{\bf Experimental results on $w(T,x)$}
\end{center}

In the previous section, we have explained how the logarithmic 
derivative $w(T,x)$, defined by Eq.\ (\ref{log_deriv}), yields valuable 
information on the MIT without the need of performing fits to a 
microscopic theory. Now we consider five materials from this point of 
view. 

A detailed study, using this approach, of amorphous Si$_{1-x}$Ni$_x$ 
\cite{MF98} was recently published, cf.\ Fig.\ 1. Within this 
investigation, sample sets prepared by two different preparation 
technologies were compared. Figs.\ 5 and 6 show part of the $w(T)$ data 
obtained in \cite{MF98}. In these graphs, as well in Figs.\ 7--10, $w$
is represented versus $T^{1/2}$ in order to stretch the low-$T$ in
comparison to the high-$T$ part. Moreover, for the samples with 
$w \ll 1/2$,  $w$ would be approximately proportional to $T^{1/2}$, if, 
as often assumed, Eq.\ (\ref{met_extrap}) with $p = 1/2$ would hold.

For both data sets presented in Figs.\ 5 and 6, at sufficiently 
high $T$, all samples studied behave very similarly: $w(T)$ increases 
with increasing $T$. On the other hand, for the samples with the 
smallest Ni content, at low $T$, there is a pronounced increase of 
$w(T)$ with decreasing $T$, indicating activated conduction. The 
strength of this contribution decreases with increasing $x$. One common 
feature of all samples (with $w > 0$), which exhibit such an increase of 
$w$ with decreasing $T$, has to be stressed: this increase is always 
found to continue down to the lowest accessible $T$. The generalisation 
of this finding is a basic assumption of our data analysis.

The differing behaviour at low and high $T$ must be caused by two
different mechanisms being dominant at low and high $T$, respectively.
The low-$T$ mechanism related to increasing $w(T)$ as $T$ decreases is
very likely a kind of hopping conduction, cf.\ 
\cite{ME83,MV85,RH97b,ZZ84}. Concerning the high-$T$ contribution 
it was speculated in \cite{MV85} that electron-phonon interaction might 
be the origin.

The minimum of $w(T)$ related to the crossover between these two 
mechanisms is particularly interesting. It is located at 
$T_{\rm min} = 150\ {\rm K}$ for the most insulating sample studied, and 
shifts to lower $T$ with increasing Ni content. Whatever the character 
of the MIT at $T = 0$, this behaviour has to be expected: if, as is 
generally taken for granted, the characteristic hopping energy tends to 
0 as the transition is approached, hopping only becomes dominant at 
lower and lower $T$. However, what provides information on the character 
of the MIT is the behaviour of the related $w_{\rm min}$. This quantity 
also seems to tend to 0 as $x \rightarrow x_{\rm c} -0$: For the samples 
1, 2, d, and e, $T_{\rm min} = 6\ {\rm K}$, 4~K, 0.8~K, and 0.2~K,
respectively, where $w_{\rm min} = 0.42$, 0.32, 0.15, and 0.06.

Figs.\ 5 and 6 also show that $w(T = {\rm const.},x)$ decreases 
monotonically as the MIT is approached from the insulating side 
(consider in particular the series a--h in Fig.\ 5). Thus 
$w(T,x_{\rm c})$ must be smaller than $w(T)$ for that insulating sample 
which is closest to the MIT. In that sense, the function 
$w_{\rm min}(T_{\rm min})$ is an upper bound of 
$w(T_{\rm min},x_{\rm c})$, and the probable simultaneous vanishing of 
$T_{\rm min}$ and $w_{\rm min}$ as $x \rightarrow x_{\rm c}-0$ imply 
that $w(T,x_{\rm c}) \rightarrow 0$ as $T \rightarrow 0$. Therefore, 
according to the previous section, the limit 
$\sigma(T \rightarrow 0,x_{\rm c})$ appears to be finite. It has the 
meaning of a minimum metallic conductivity because 
$\sigma(T = {\rm const.},x)$ monotonically increases with $x$. Thus the 
MIT is very likely discontinuous at $T = 0$. For a more detailed 
discussion see \cite{MF98}.

It might be worth mentioning a technical point in connection with Figs.\ 
5 and 6. The comparatively small random scatter of these $w(T)$ data 
arises not only from a high accuracy of the $R(T)$ measurements, but 
also from the numerical method used for calculating $w$. If we consider 
$\ln \sigma$ as function of $\ln T$, its slow variation permits a 
piecewise approximation of high accuracy as a second-order polynomial in 
$\ln T$. Thus, in \cite{MF98}, a window taking in a certain number, $k$, 
of neighbouring points was moved along the $\ln \sigma(\ln T)$ curve, and 
$w$ was calculated by means of linear regression from the data within 
the window. The trick is to relate the obtained slope to that $\ln T$ 
value for which the derivative of a second-order polynomial is exactly 
reproduced, independently of the values of its coefficients,
\begin{equation}
\ln T_{\rm lf} = \frac
{k \sum_i (\ln T_i)^3 - \sum_i \ln T_i \sum_j (\ln T_j)^2}
{2 (k \sum_i (\ln T_i)^2 - (\sum_i \ln T_i)^2)}\,,
\end{equation}
where the sums run over all points within the window considered. In this 
way, the total error (numerical plus random) can be kept small. Finally, 
if a set of $w(T)$ data has been calculated by simply using pairs of 
neighbouring $\sigma(T)$ points, the above procedure can be emulated by 
smoothing with appropriately chosen weights. This equivalence will be 
utilised below in analysing $w(T)$ data from the literature.

We turn now to comparison with other alloys. Amorphous Si$_{1-x}$Cr$_x$ 
prepared by electron-beam evaporation was studied in
\cite{ME83,MV85,M87,M90a,M90b}. Fig.\ 7 shows $w(T)$ for four samples,
obtained from $\sigma(T)$ data published in \cite{ME83,MV85}. Below
roughly 20~K, activated conduction is obvious for three of them from the
behaviour of $\mbox{d} w / \mbox{d} T$. Again, there are minima in the 
related $w(T)$, and these $w(T)$ take also values which are far smaller 
than 1/2.

Recent results on amorphous Si$_{1-x}$Mn$_x$ by Yakimov et al.\ (see 
Fig.\ 4 of \cite{YD98}), corroborate the qualitative behaviour of 
$w(T,x)$ which we have presented for amorphous Si$_{1-x}$Ni$_x$ and 
Si$_{1-x}$Cr$_x$. In Fig.\ 8 here, we show $w(T)$ relations obtained by 
smoothing the original Si$_{1-x}$Mn$_x$ data according to the procedure 
mentioned above. 

From the examples presented, the question may arise as to whether the 
existence of low-$T$ pieces of $w(T)$ curves with simultaneously 
$w \ll 1/2$ and $\mbox{d} w / \mbox{d} T \le 0$ -- very likely caused by 
a discontinuity of $\sigma(0,x)$ at $x_{\rm c}$ -- is specific for 
(some) alloys of semiconductors with transition metals only. That this 
is not the case is shown by the following two examples. Fig.\ 9 presents 
two $w(T)$ curves for samples of amorphous InO$_x$ (data from Figs.\ 8 
and 9 of \cite{GA98}). These samples seem to exhibit activated 
conduction although $w$ is very small, i.e., of the order of 0.01. Thus 
the carrier concentration must be close to its critical value in both 
cases. Fig.\ 10 shows $w(T)$ curves for three Ge-Al samples (data from 
Fig.\ 2 of \cite{RS94}). Although these samples do not have an amorphous 
structure, but consist of small 
($\stackrel{_{\textstyle <}}{_{\textstyle \sim}} 20\ {\rm \AA}$) Al 
grains embedded in an 
amorphous matrix, $w(T,x)$ exhibits the typical features found with the 
amorphous materials studied above. Thus clustering seems not to destroy 
this qualitative behaviour, though it may change the value of the
minimum metallic conductivity.

To summarise, for all the five materials considered, the qualitative
behaviour of $w(T,x)$ does not appear to be consistent with the usual 
assumption that the MIT is continuous at $T = 0$, and that 
$\sigma(T,x_{\rm c}) \propto T^p$ with $p = 1/2$ or $1/3$. It suggests 
that $w(T,x_{\rm c})$ tends to 0 as $T \rightarrow 0$, in which case, 
according to the arguments presented earlier, the MIT should be 
discontinuous at $T = 0$ (but only there). 

\begin{center} \noindent
{\bf Phenomenological model of $\sigma(T,x)$}
\end{center}

The analysis in the previous section has illuminated a qualitative 
feature of the MIT in amorphous alloys: for several materials, it seems 
likely that there is a finite minimum metallic conductivity. However, 
for identification of physical mechanisms, quantitative descriptions
are needed, where homogenous systems have first of all to be understood. 
Highly stable amorphous alloys, in which the formation of crystalline 
regions proceeds exceptionally slowly, are particularly suitable for 
this. They offer the best chances for observing the corresponding 
dependences in an almost clean form. Such a substance is amorphous 
Si$_{1-x}$Cr$_x$, which is used in microelectronics as a basis material 
of thin-film resistors \cite{HS93}.

Amorphous Si$_{1-x}$Cr$_x$ films prepared by electron-beam 
evaporation have been investigated in \cite{ME83,MV85,M87,M90a,M90b}. A 
particularly important result of these studies is the detection of a 
scaling law for the $T$ dependence in the hopping region. For 
$T \stackrel{_{\textstyle <}}{_{\textstyle \sim}} 20\ {\rm K}$, the 
conductivity can be described by
\begin{equation}
\sigma(T,x) = \sigma_0 \cdot \varphi(T/T_0(x))\,.
\label{hop_scal}
\end{equation}
This universality was detected by two ansatz-free methods: 
(i)~graphical construction of a master curve \cite{ME83}, see Fig.\ 11, 
and (ii)~proving that $w = \mbox{d} \ln \sigma / \mbox{d} \ln T$ is 
completely determined merely by $\ln \sigma$ (see Fig.\ 2 in 
\cite{MV85}), i.e., by demonstrating that, in a $w$ versus $\ln \sigma$ 
plot, the data points from different samples form a common curve. The 
validity of the scaling law (\ref{hop_scal}) has an important physical 
meaning: it shows that a single mechanism governs the conduction. Note, 
that this general result was obtained without performing fits of an 
ansatz based on microscopic theory. 

For high temperatures, i.e., as $T/T_0 \rightarrow \infty$, the 
function $\varphi(T/T_0)$ seems to saturate \cite{MV85}. Therefore,
it is suggested to eliminate the ambiguity of Eq.\ (\ref{hop_scal}) 
concerning the value of $\sigma_0$ by defining $\varphi(\infty) =1$.
On the other hand, in the exponential region, Eq.~(\ref{hop_scal}) takes 
the form  of Eq.\ (\ref{exp_hopping}) with $\nu \approx 0.5$ -- an 
indication of the relevance of electron-electron interaction -- see
\cite{ME83,MV85}. It follows from the scaling law that the related 
pre-factor $\sigma_1$ is independent of $x$. This seems to be the case 
also in several other substances \cite{M85}, cf.\ \cite{ZZ84}. It is 
remarkable that $\sigma_1$ is by a factor of 2.5 smaller than 
$\sigma_0$ \cite{MV85}.

This scaling behaviour is broken in two ways. On the one hand,
increasing the temperature above roughly 20~K (the specific value 
depends on the accuracy demanded) causes a faster increase of $\sigma$
than expected according to extrapolation by means Eq.~(\ref{hop_scal}).
It is remarkable that this feature can be described by a multiplicative 
decomposition,
\begin{equation}
\sigma(T,x) = \sigma_0 \cdot \varphi(T/T_0(x)) \cdot \xi(T,x)\,,
\label{ph_mod_sc}
\end{equation}
with $\xi(0,x) = 1$: In \cite{MV85}, it was observed that the high-$T$ 
factor $\xi(T,x)$ is almost independent of $x$ close to the MIT, and 
that it depends on $T$ almost linearly between roughly 100 and 300~K. 
Given this independence of $x$, the high-$T$ deviations from the scaling 
law (\ref{hop_scal}) probably originate from a single mechanism, which, 
due to the continuity of $\sigma(T = {\rm const.},x)$ at $x_{\rm c}$ for 
$T > 0$, seems to `survive' the MIT.  The nature of this mechanism is 
unclear as yet; it was speculated in \cite{MV85} that electron-phonon 
interaction could play an important role. On the other hand, annealing 
the samples causes changes of the pre-factor $\sigma_1$ in the hopping 
law (\ref{exp_hopping}) \cite{MV85}, and, for sufficiently high 
annealing temperature, even the value of the exponent $\nu$ seems to be 
shifted. This arises presumably from a second length scale becoming 
relevant, namely the size of CrSi$_2$ clusters formed \cite{SB91}.

Also, in the metallic region, a high-$T$ mechanism (presumably the same 
as on the activated side) plays a substantial role. Here, it competes 
with a low-$T$ mechanism which causes the opposite behaviour, i.e., an 
increase of $\sigma$ with decreasing $T$. In \cite{M90a}, the detailed 
study of this low-$T$ contribution was made possible by a trick: 
Starting from the observation that, in a related graph, the $\sigma(T)$ 
curves of the individual samples are almost parallel to each other at 
high $T$, the difference between the $\sigma$ values of the sample under 
consideration and a reference sample was calculated. Assuming additivity 
of the $\sigma$ contributions related to the different mechanisms, this 
procedure considerably reduces the influence of the high-$T$ mechanism, 
and thus magnifies the low-$T$ contribution. Therefore the latter could 
be studied within a broader $T$ interval. A series of power-law fits of 
this difference were performed in \cite{M90a} to determine the exponent 
$p$ in Eq.\ (\ref{met_extrap}). In doing so, unphysical dependences of 
the $p$ value on the bounds of the $T$ and $x$ regions considered were 
carefully excluded. The fits which simultaneously took into account 15 
samples yielded $p = 0.19 \pm 0.03$. Finally, in \cite{M90b}, the 
analysis of the correlations between the parameters $a$ and $b$ of the 
approximation (\ref{met_extrap}) lead to the following empirical 
description of $\sigma(T,x)$ in the metallic region:
\begin{equation}
\sigma(T,x) = a(x) + b(x) \cdot T^p + \sigma_0 \cdot \xi(T,x)\,,
\label{ph_mod_me}
\end{equation}
where $b \le 0$. The relation between the parameters $a$ and $b$ 
exhibits the singularity
\begin{equation}
b \propto a^\vartheta
\end{equation}
with the critical exponent $\vartheta = 0.68 \pm 0.05 - (p-0.19)$. Since
$a^\vartheta$ is not defined for $a < 0$, this singularity can be 
considered as another independent indication of the MIT. 

Equations (\ref{ph_mod_sc}) and (\ref{ph_mod_me}) together form a simple 
phenomenological model of $\sigma(T,x)$ close to the MIT. Guaranteeing 
continuity at $x_{\rm c}$ for $T > 0$, they describe both sides of the 
transition. For further details, in particular small low-$T$ deviations 
and consequences from this model, we refer to \cite{M90b}.

\begin{center} \noindent
{\bf Conclusions}
\end{center}

Summarising, we have analysed $\sigma(T,x)$ data from several amorphous 
alloys close to the MIT in an as unbiased way as possible, avoiding the 
use of results from current microscopic theories to a large extent. For 
that, we have considered the $T$ dependence of the logarithmic 
temperature derivative of the conductivity,
$w(T,x) = \mbox{d} \ln \sigma / \mbox{d} \ln T$. In various cases,
the behaviour of this quantity differs qualitatively from the 
predictions obtained from commonly accepted theory. There are samples, 
for which $w$ simultaneously takes values far smaller than $1/2$, and 
then increases with decreasing $T$ down to the lowest temperatures 
studied. Hence, it is suggested that $w(T,x_{\rm c})$ tends to 0 as 
$T \rightarrow 0$. This limiting behaviour indicates a discontinuity of 
$\sigma(0,x)$, so that Mott's hypothesis of the existence of a finite 
minimum metallic conductivity is probably correct in these systems after 
all. However, it does not imply a discontinuity of 
$\sigma(T={\rm const.},x)$ at the MIT for any $T > 0$, neither does it
permit any conclusion as regards the validity of the arguments Mott used 
to estimate a value for minimum metallic conductivity. Hence, we believe 
that, for amorphous alloys, the `critical-exponent puzzle' concerns only 
an artifact arising from unphysical fitting.

For amorphous  Si$_{1-x}$Cr$_x$, a comparatively stable alloy, one
can go on, and derive a quantitative phenomenological model by searching 
for universal features in $\sigma(T,x)$ \cite{ME83,MV85,M87,M90a,M90b}. 
On the activated side of the MIT, a scaling law for the temperature 
dependence, $\sigma(T,x) = \sigma_0 \cdot \varphi(T/T_0(x))$, is the 
basis of this model. However, annealing destroys the scaling behaviour 
so that a highly amorphous structure seems to be the necessary 
prerequisite for this universality. On the metallic side, a negative 
$\sigma(T)$ contribution proportional to $T^p$ with $p = 0.19 \pm 0.03$ 
governs the low-temperature behaviour. The related proportionality 
factor vanishes as the MIT is approached. It is remarkable that, in the 
exponential region, the above scaling law can also be observed in 
several other substances \cite{ZZ84,M85}.

Comparing the characteristic features of this phenomenological model
with basic assumptions of and conclusions from current microscopic
theories permits speculations on the physical nature of the MIT in
these systems, see \cite{M85,M86}. It is particularly important that the 
model yields $\mbox{d} \sigma / \mbox{d} T = 0$ at $x_{\rm c}$ for 
sufficiently low $T$. According to \cite{M85,M86}, the MIT cannot be 
understood in terms of an Anderson transition, but it should be 
caused by the breakdown of screening when the MIT is approached 
from the metallic side. This should be accompanied by the formation of a 
Coulomb gap having the width 0 at $x_{\rm c}$ and opening when moving 
into the insulating $x$ region. Our speculation on the nature of the MIT 
coincides with the microscopic theory very recently published by Chitra 
and Kotliar \cite{CK99}. These authors incorporate the long-range 
Coulomb interaction into dynamical mean-field theory, and obtain that 
the MIT should be discontinuous in two- and three-dimensional systems.

In conclusion, the review given above should not be understood as a
`final' solution concerning the character of the MIT in amorphous 
alloys. Instead, our aims have been to demonstrate by an unbiased 
analysis that this topic cannot be considered as closed, and to 
stimulate further research. There are several interesting questions for
future experiments: (i)~How does $w(T)$ behave for other alloys? Is the 
existence of samples with simultaneously $w \ll 1/2$ and 
$\mbox{d} w / \mbox{d} T < 0$ being valid down to the lowest measuring 
temperatures indeed a general feature? (ii)~Following \cite{AM96}, 
under which conditions does one observe metallic $\sigma(T)$ curves with 
$\mbox{d} \sigma / \mbox{d} T < 0$? This means what influence have 
composition, magnetic moments, structure, and preparation conditions on 
this phenomenon, and on the bounds of the $T$ interval where this 
mechanism is dominant? (iii)~Can the phenomenological model, which was 
obtained for amorphous Si$_{1-x}$Cr$_x$, also be used for the 
description of other comparatively stable amorphous alloys? In 
particular, is the scaling function $\varphi(T/T_0(x))$ for the 
non-exponential region substance independent as, provided there is no 
clustering, it seems to be in the exponential region \cite{M85}? What 
about the $T$ exponent in the metallic region? (iv)~Can this model be 
extended to the description of other quantities such as the 
magnetoresistance? (v)~Last, but not least, it should be noted that
the conclusions drawn above are very different from the description of 
heavily doped crystalline semiconductors given in a previous review 
in this journal, i.e., in \cite{L98}. On the other hand, Fig.\ 1 of 
\cite{HT89} and Fig.\ 6 of \cite{SK89} show for crystalline Si:(P,B) and 
Si:As, respectively, that the behaviour of $w(T)$ resembles the results 
for amorphous alloys. How can these contradictions be resolved?
\\

\begin{center}
\noindent
{\bf Acknowledgements}
\end{center}

This review is based on the studies 
\cite{ME83,MV85,M87,M90a,M90b,YD98,MF98,RS94,GA98}. We thank all our 
colleagues who contributed to these publications. In particular, 
critical and stimulating discussions with R.~Rosenbaum, M.~Schreiber, 
and A.I.~Yakimov are gratefully acknowledged. Moreover, we are obliged 
to A.I.~Yakimov for sending us tables of the data presented in Fig.\ 4 
of \cite{YD98}.

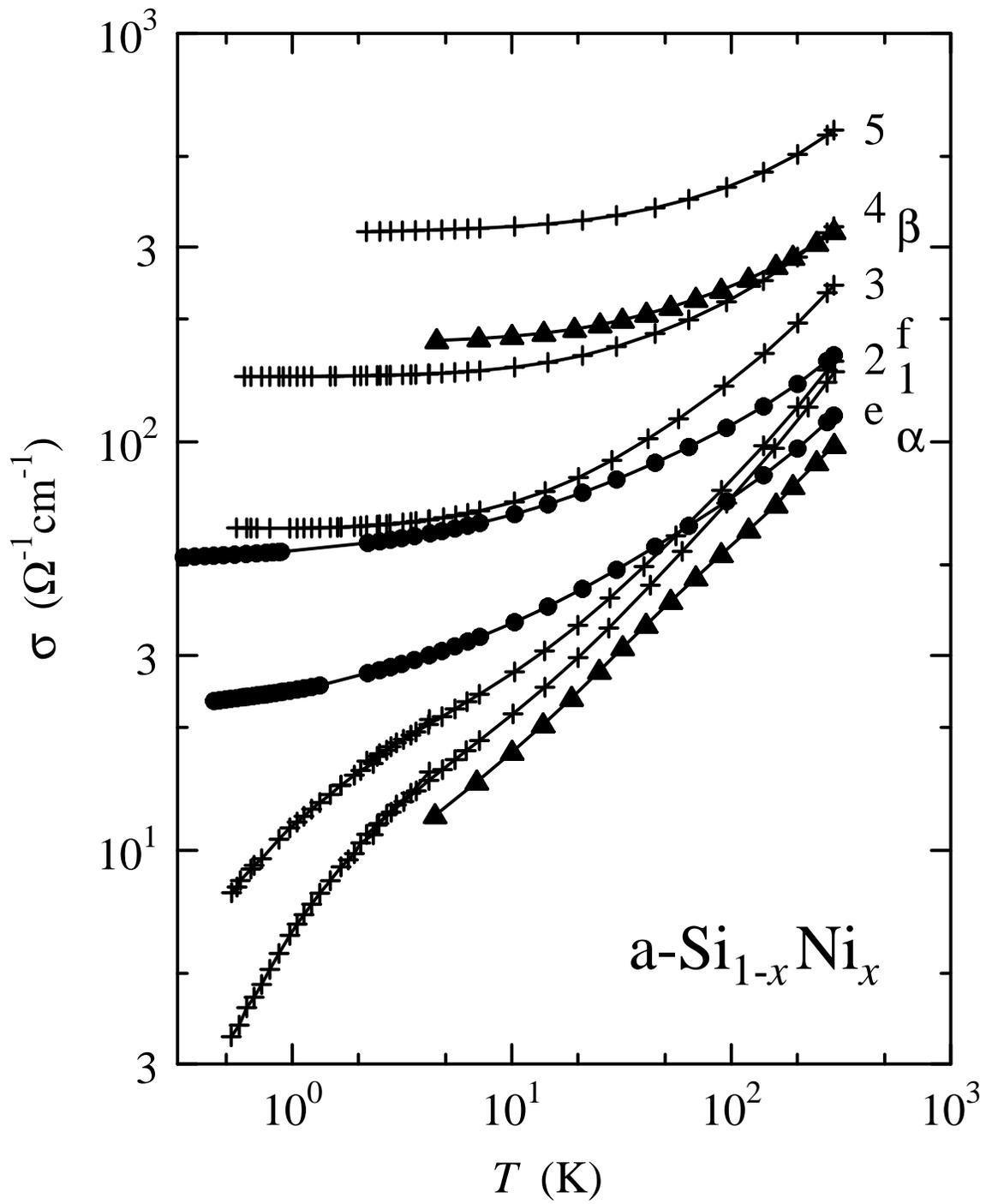
\begin{figure}
\caption{
Overview of $\sigma(T,x)$ for a-Si$_{1-x}$Ni$_x$ based on data from 
{\protect \cite{MF98,IY97}}. The samples 1--5 (+) were prepared by 
electron-beam evaporation of the alloy {\protect \cite{MF98}}, the 
samples e and f ($\bullet$) were deposited by a gradient technique 
based on co-evaporation of the elements {\protect \cite{MF98}}, and the 
samples $\alpha$ and $\beta$ ($\blacktriangle$) were sputtered 
{\protect \cite{IY97}}. Ni content: $x = 0.149$, 0.164, 0.167, 0.189, 
0.221 (RBS) for samples 1--5, respectively; $x = 0.235$, 0.248 (EDX, 
these values exceed related RBS data by roughly 0.06 
{\protect \cite{MF98}}) for samples e and f, respectively; $x = 0.207$, 
0.271 (EDX) for samples $\alpha$ and $\beta$, respectively.
}
\label{fig1}
\end{figure}

\begin{figure}
\caption{
Experimental parameter plane: The insulating region, where only
activated conduction occurs, is marked by I, the metallic region by M,
and the lowest experimentally accessible temperature by $T_{\rm lea}$.
The characteristic hopping temperature, $T_0(x)$, is represented by a 
full line. Measuring $\sigma(T,x = {\rm const.})$ means to obtain data 
points ($\bullet$, $\times$) along vertical lines. Only for $T < T_0$ 
does $\sigma$ depend exponentially on $T$. For $T > T_0$, comparatively 
flat, non-exponential $\sigma(T)$ dependences are expected. Thus, for 
$\bullet$, depending on $T$, both exponential and non-exponential 
behaviour is observed. However, for $\times$, only non-exponential 
$\sigma(T,x = {\rm const.})$ is found, although this sample belongs to 
the insulating region, too. The latter problem occurs in the whole 
interval $[x^\star,x_{\rm c})$. (Slightly modified reproduction of
Fig.\ 1 from \protect\cite{MF98}.)
}
\label{fig2}
\end{figure}

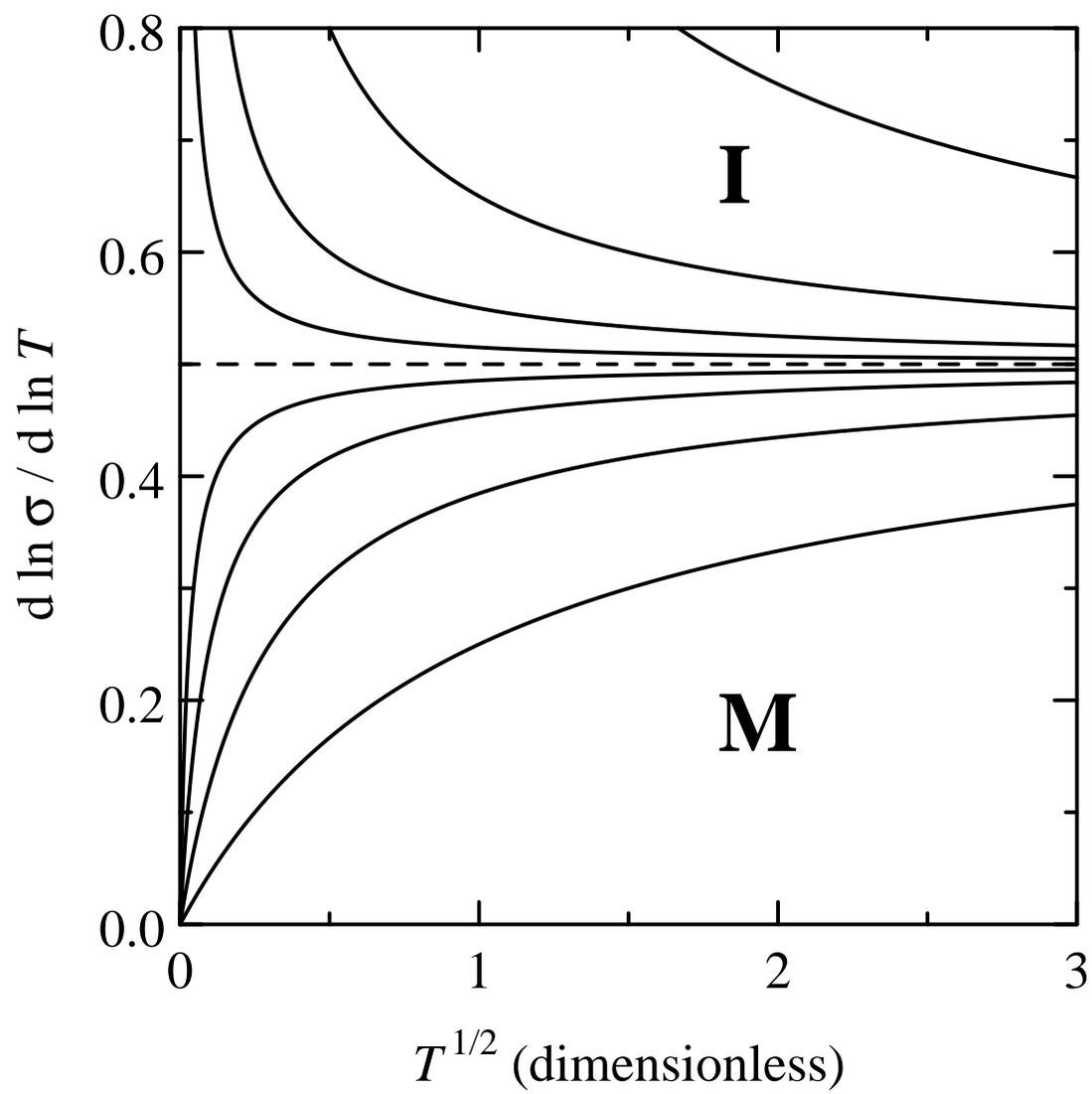
\begin{figure}
\caption{
Logarithmic derivative $w(T,x) = \mbox{d} \ln \sigma / \mbox{d} \ln T$ 
for a continuous transition according to Eq.\ (6). The (at $T = 0$) 
insulating region is marked by I, the metallic region by M. The dashed
line represents the separatrix between these areas. The full lines 
correspond to the following parameter values: $T_0(x)^{1/2} = 0.03$, 
0.1, 0.3, 1 in the insulating region, and $a(x) = 0.03$, 0.1, 0.3, 1 in
the metallic area.
}
\label{fig3}
\end{figure}

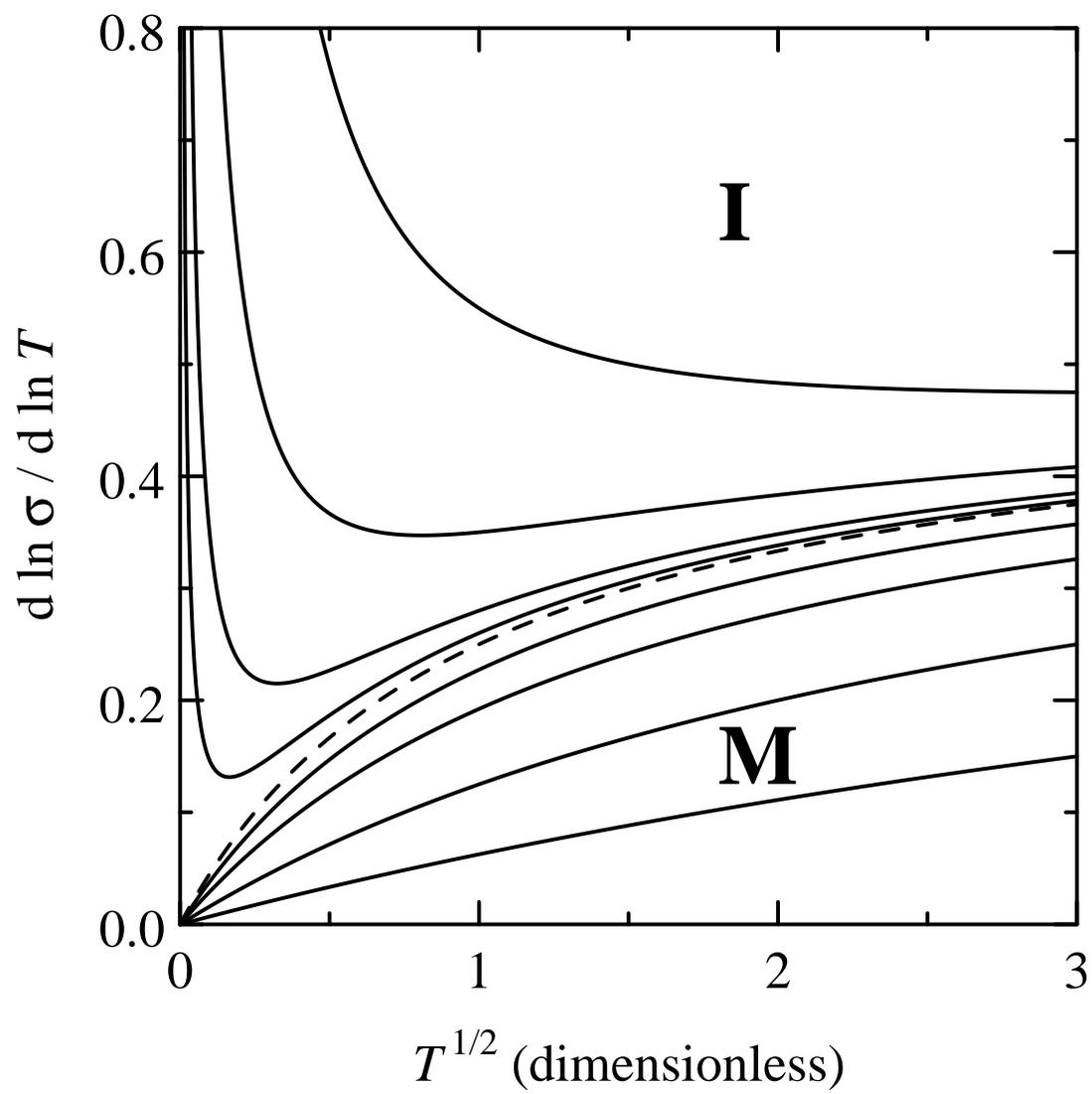
\begin{figure}
\caption{
Logarithmic derivative $w(T,x) = \mbox{d} \ln \sigma / \mbox{d} \ln T$
for a discontinuous transition according to Eq.\ (7). The full lines 
correspond to the following parameter values: $T_0(x)^{1/2} = 0.02$, 
0.06, 0.2, 0.6 in the insulating region, and $a(x) - 1 = 0.2$, 0.6, 2, 6 
in the metallic area. For further details see caption of Fig.\ 3.
}
\label{fig4}
\end{figure}

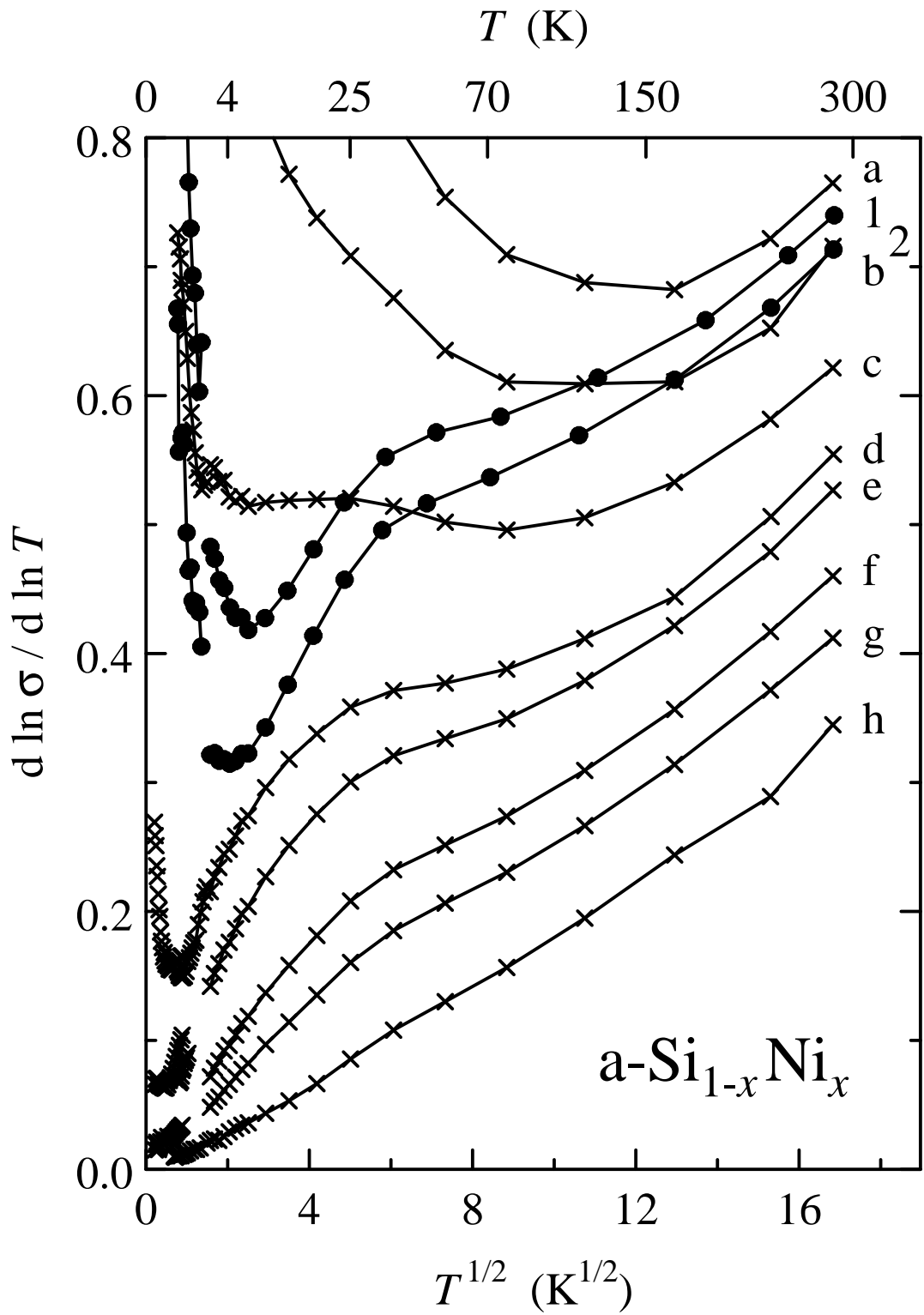
\begin{figure}
\caption{
$w(T,x) = \mbox{d} \ln \sigma / \mbox{d} \ln T$ for a-Si$_{1-x}$Ni$_x$,
presenting data from Figs.\ 7a and 7b of {\protect \cite{MF98}}. 
Ni content: $x = 0.149$, 0.164 (RBS) for samples 1 and 2, respectively; 
$x = 0.195$, 0.203, 0.212, 0.223, 0.235, 0.248, 0.264, 0.282 (EDX, 
these values exceed related RBS data by roughly 0.06 
{\protect \cite{MF98}}) for samples a--h, respectively.
}
\label{fig5}
\end{figure}

\begin{figure}
\caption{
Low-$T$ / low-$w$ part of Fig.\ 5. The data presented were obtained from
measurements in three different cryostats, where several months passed 
between the investigations. Samples g and h are metallic films
\protect\cite{MF98}, included for comparison; for sample f, a conclusion 
on the character of the conduction could not be drawn within 
\protect\cite{MF98}. (Modified reproduction of Fig.\ 8b from 
\protect\cite{MF98}.)
}
\label{fig6}
\end{figure}

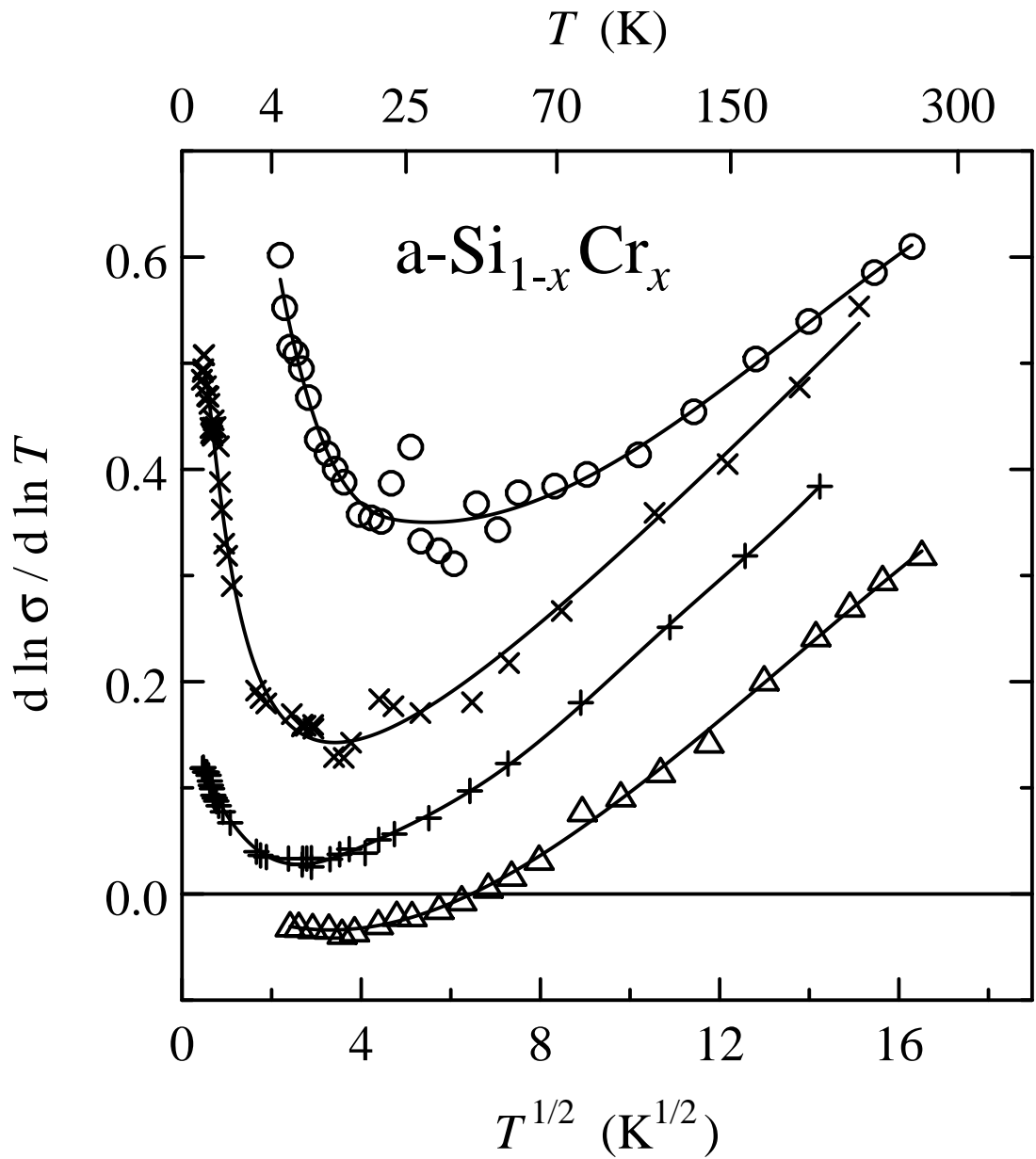
\begin{figure}
\caption{
$w(T,x) = \mbox{d} \ln \sigma / \mbox{d} \ln T$ for a-Si$_{1-x}$Cr$_x$,
obtained from data published in {\protect \cite{ME83,MV85}} (mainly
given in Fig.\ 3 of {\protect \cite{MV85}}). Cr content: $x = 0.109$ 
($\circ$), 0.127 ($\times$), 0.149 ($+$), and 0.193 ($\bigtriangleup$).
The sample marked by $\bigtriangleup$ is a metallic one included for
comparison.
}
\label{fig7}
\end{figure}

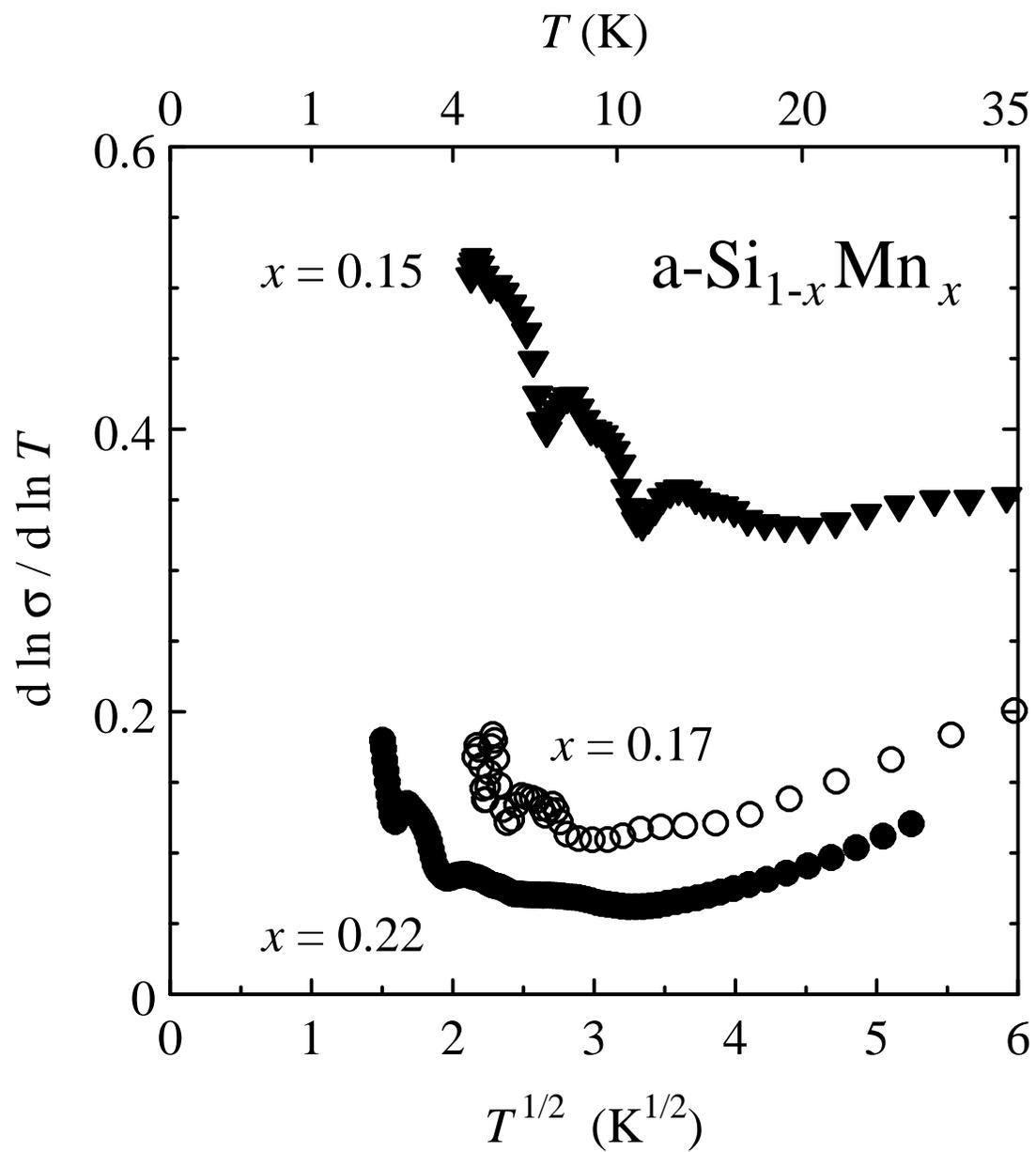
\begin{figure}
\caption{
$w(T,x) = \mbox{d} \ln \sigma / \mbox{d} \ln T$ for a-Si$_{1-x}$Mn$_x$,
obtained from the $w(T,x)$ data published in Fig.\ 4 of 
{\protect \cite{YD98}}. The $T$ dependences are smoothed here, see text, 
using windows of 15, 15, and 45 points for $x = 0.15$, 0.17, and 0.22, 
respectively.
}
\label{fig8}
\end{figure}

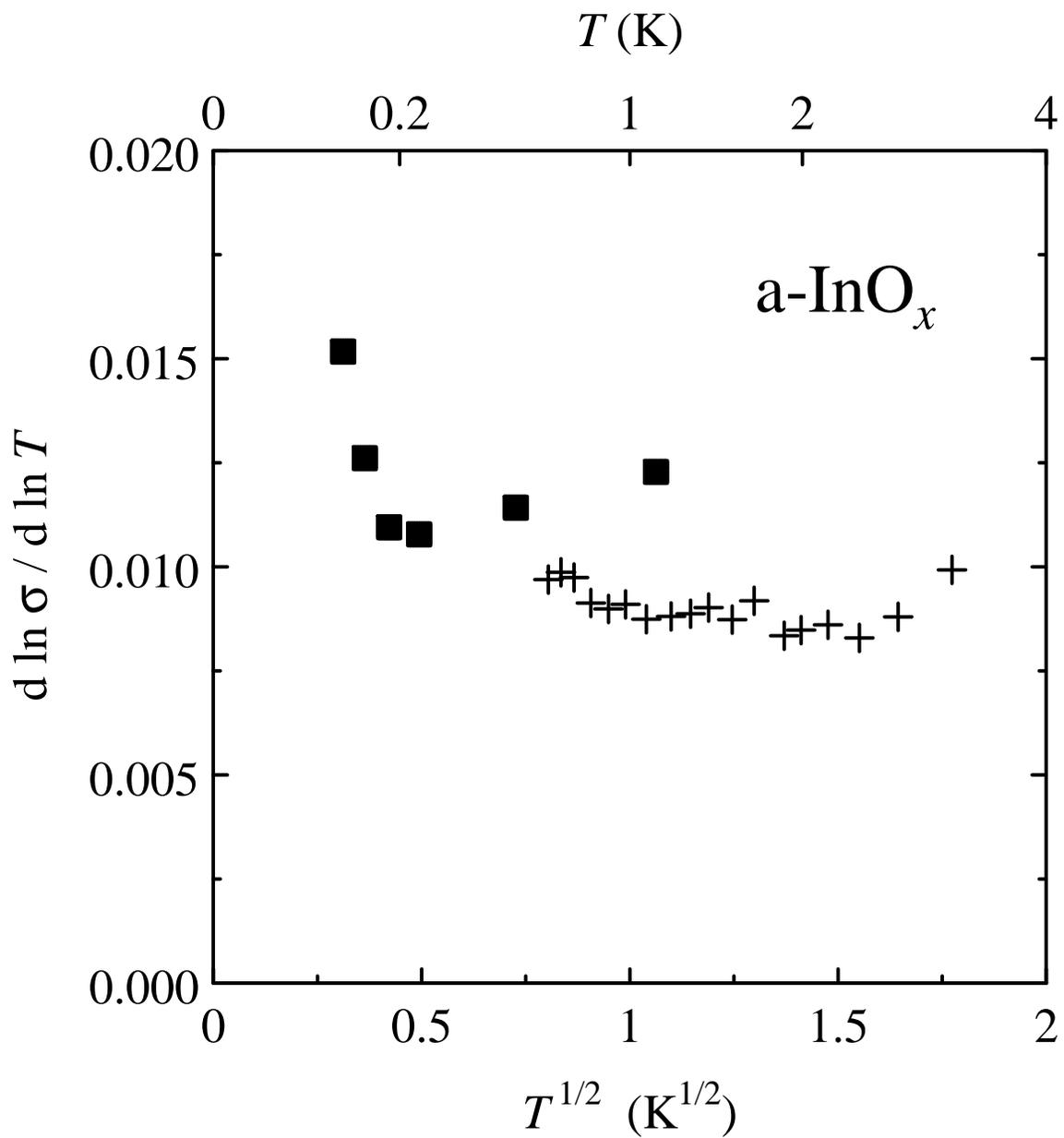
\begin{figure}
\caption{
$w(T,x) = \mbox{d} \ln \sigma / \mbox{d} \ln T$ for a-InO$_x$, obtained
from data published in Figs.\ 8 ($\sigma(T)$ for sample 3) and 9 ($w(T)$ 
for sample 42) of {\protect \cite{GA98}}. Here, $w(T)$ was calculated 
using windows of 4 and 5 points for sample 3 (+) and sample 42 
($\blacksquare$), respectively.
}
\label{fig9}
\end{figure}

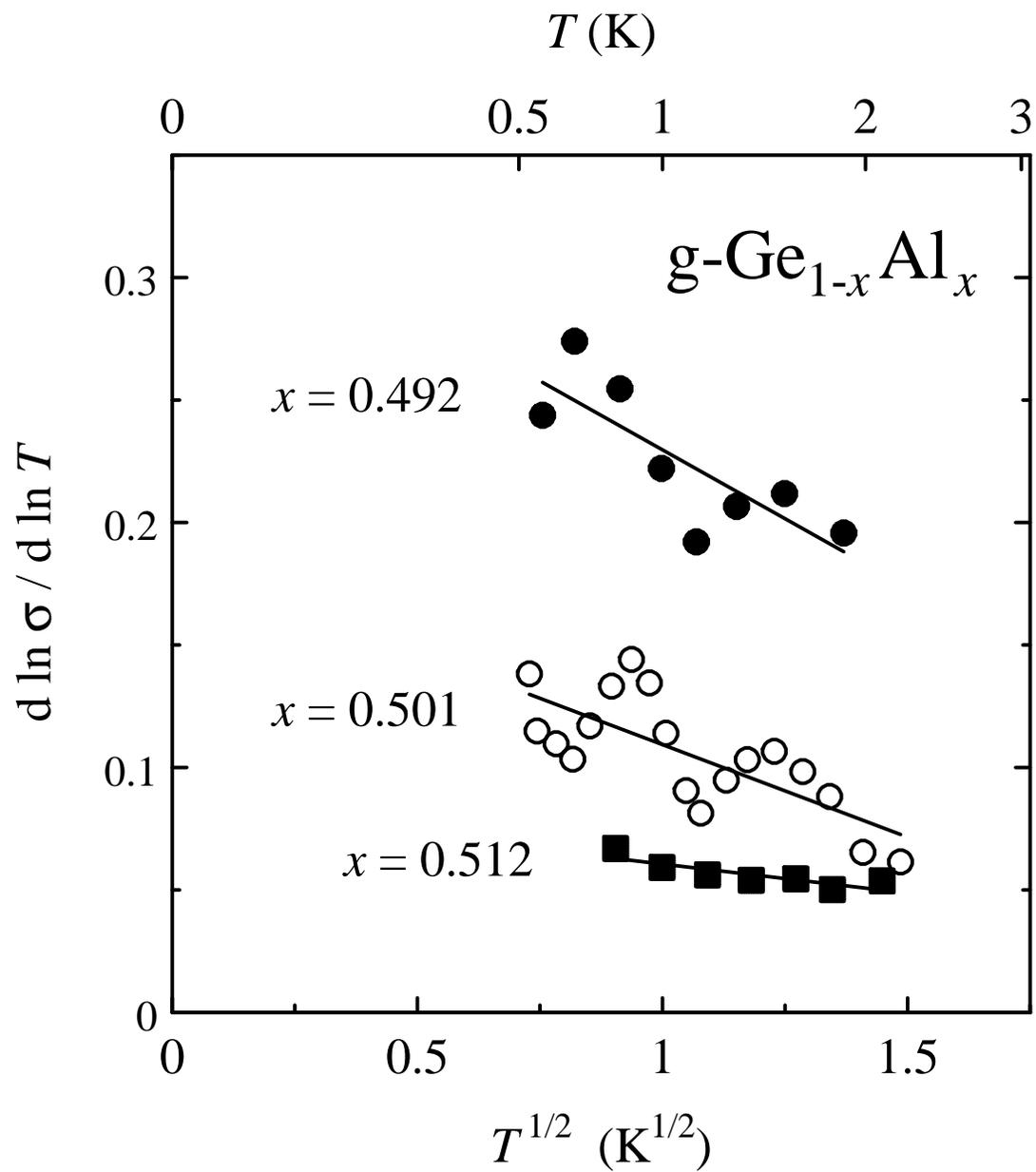
\begin{figure}
\caption{
$w(T,x) = \mbox{d} \ln \sigma / \mbox{d} \ln T$ for granular 
Ge$_{1-x}$Al$_x$, based on $w(T,x)$ data from Fig.\ 2 of
{\protect \cite{RS94}}. The corresponding $R(T)$ measurements were
performed in a magnetic field of 3.5~T to suppress superconductivity 
effects. Here, $w(T)$ was smoothed using windows of 3 points.
}
\label{fig10}
\end{figure}

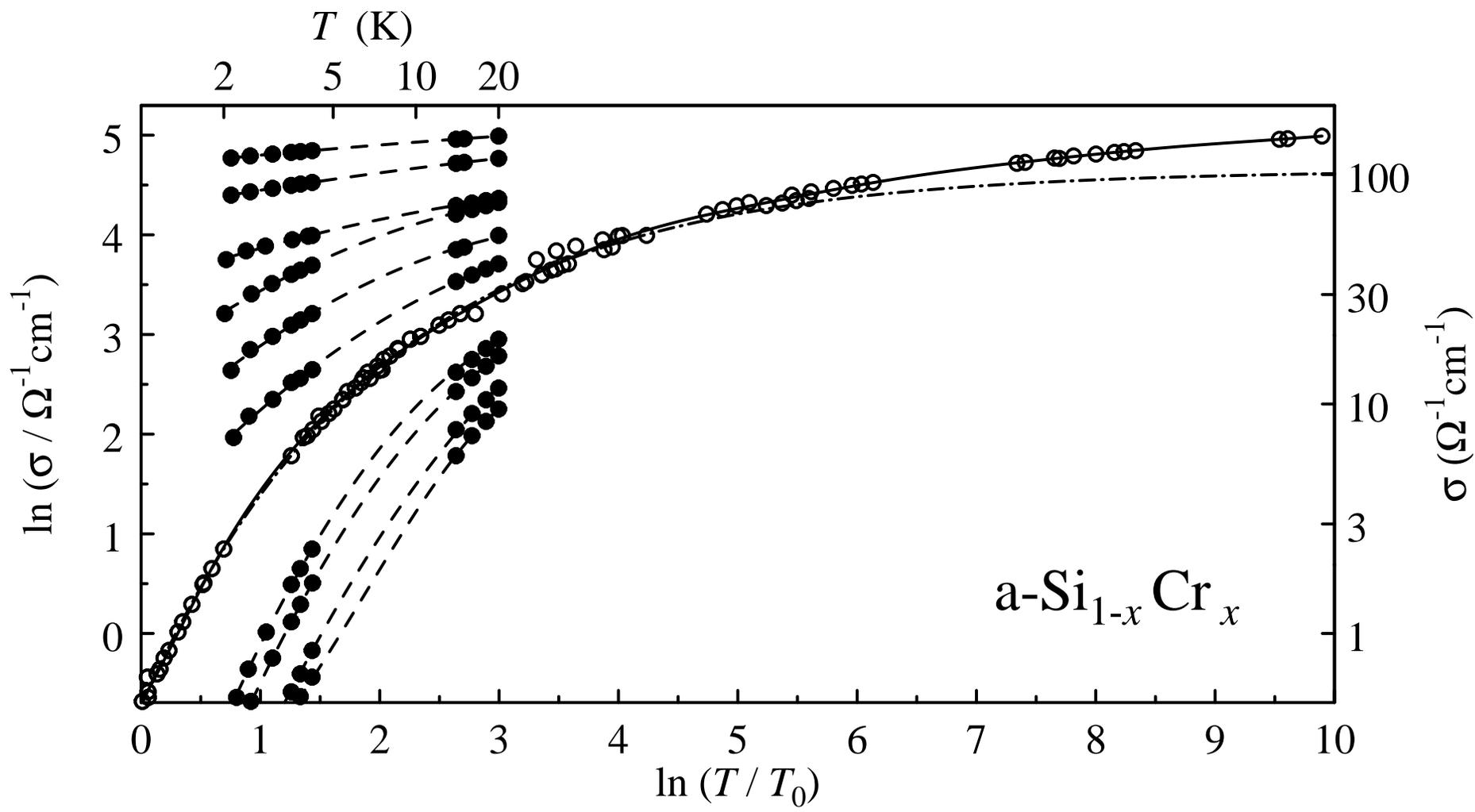
\begin{figure}
\caption{
Test of scaling behaviour of $\sigma(T,x)$ for a-Si$_{1-x}$Cr$_x$
by means of the construction of the master curve (full line) of the 
$\sigma(T)$ relations (dashed) for ten samples, taken from Fig.\ 4 of 
{\protect \cite{ME83}}. (For one of the samples, a thickness inaccuracy 
is corrected here.) The dashed-dotted curve represents the asymptotic 
exponential behaviour, $\sigma \propto \exp[-A\,(T_0/T)^{1/2}]$ (the 
constant $A$ arises from the arbitrariness in fixing the $T$ scale of
the mastercurve).
}
\label{fig11}
\end{figure}

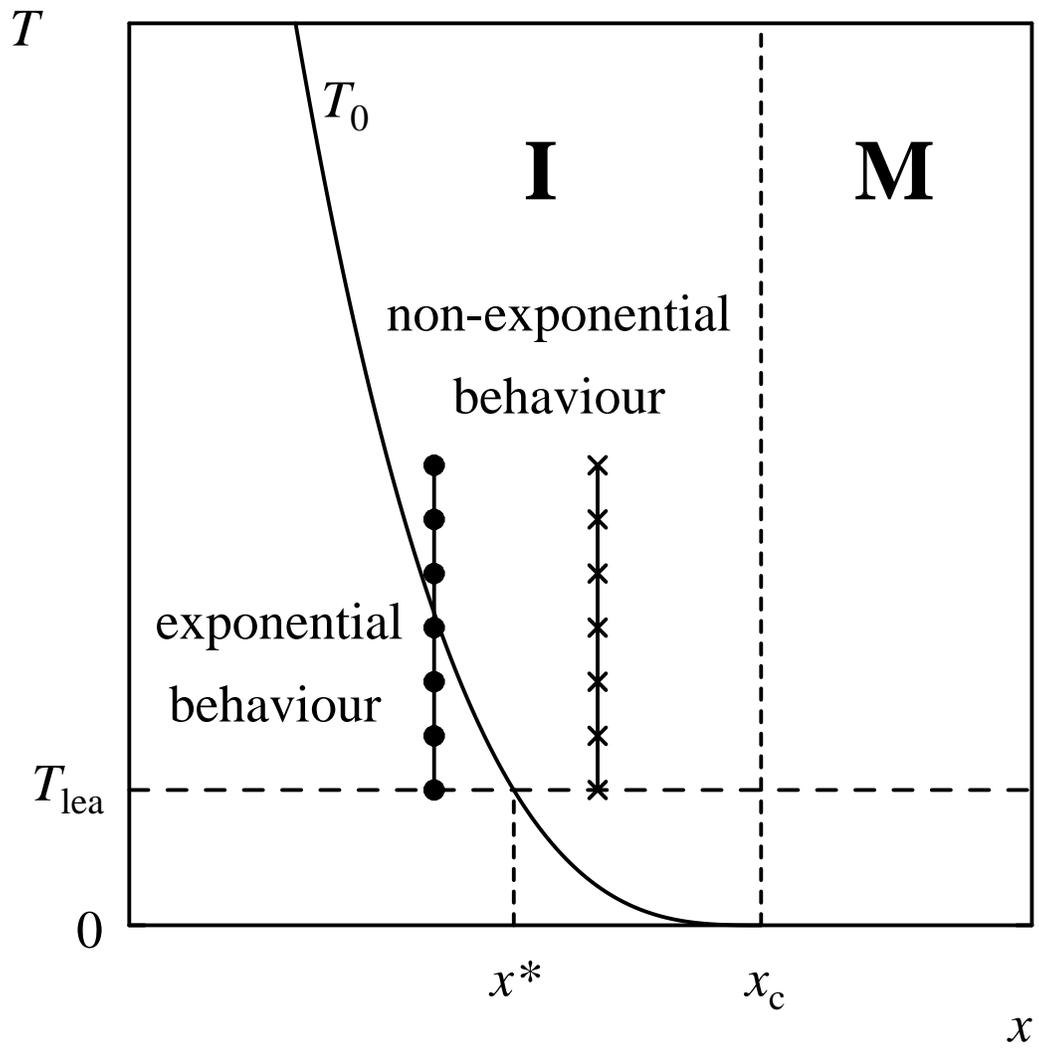
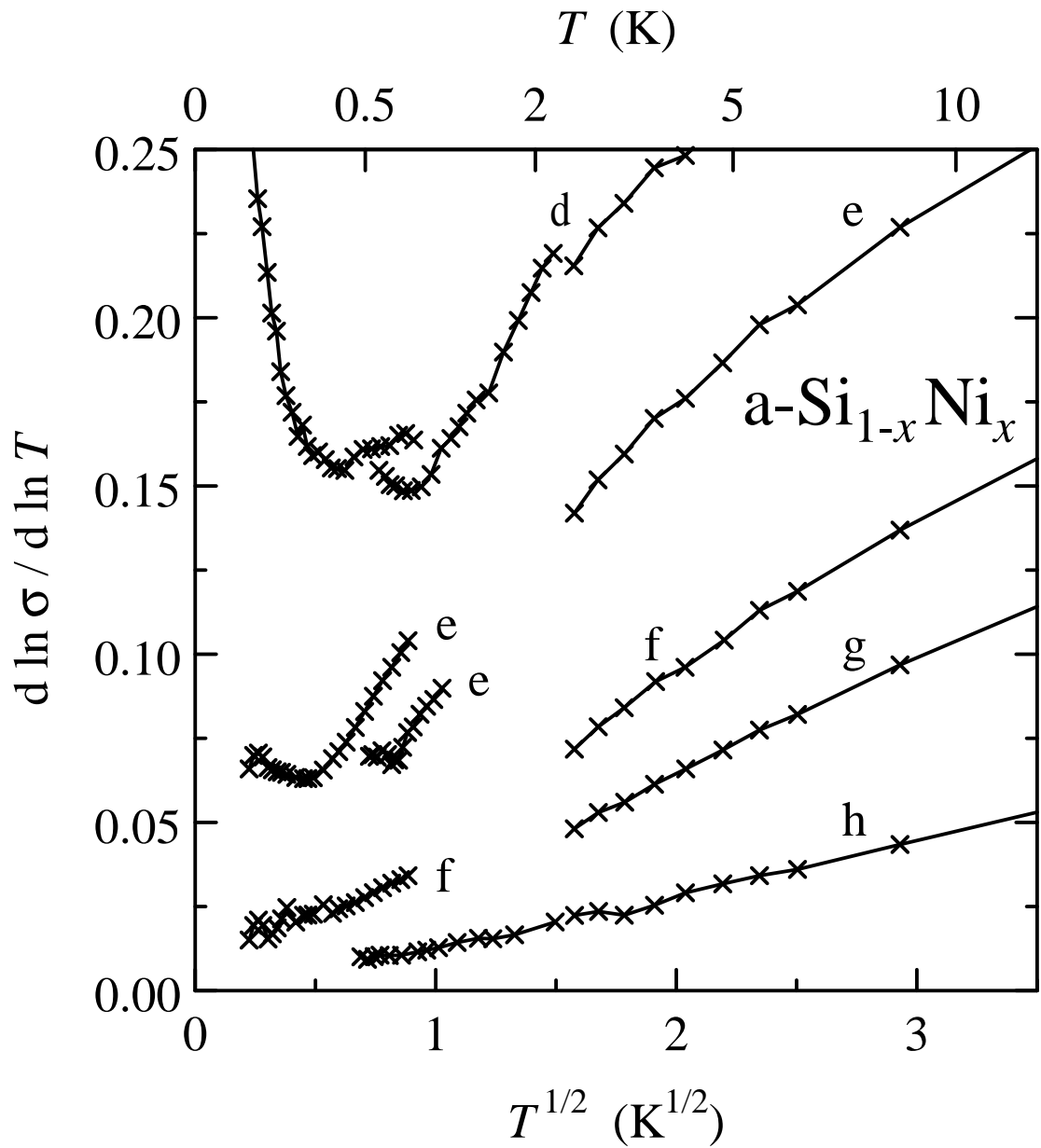
\end{document}